\documentclass[aps,prb,twocolumn,floatfix,superscriptaddress]{revtex4-1}
\usepackage{times}
\usepackage{epsfig}
\usepackage{amsmath}
\usepackage{amssymb}
\usepackage{graphicx}
\usepackage{bbold}
\usepackage{hyperref}
\usepackage[switch]{lineno}
\usepackage{pifont}
\usepackage{bbding}
\usepackage{array}
\usepackage{multirow}
\hypersetup{colorlinks=true,citecolor=red,linkcolor=blue}

\newcommand{\tsh}{t_\mathrm{h}}

\begin{document}
\title{Laser-Induced Commensurate-Incommensurate Transition of Charge Order \\
in a Hubbard Superlattice}
\author{Hua Chai}
\affiliation{College of Physics and Technology, Guangxi Normal University, Guilin, 541004, China}
\author{Zhenyu Cheng}
\affiliation{College of Physics and Technology, Guangxi Normal University, Guilin, 541004, China}
\author{Qinxin Hu}
\affiliation{College of Physics and Technology, Guangxi Normal University, Guilin, 541004, China}
\author{Zhongbing Huang}
\affiliation{School of Physics, Hubei University, Wuhan 430062, China}
\author{Xiang Hu}
\email{Contact author: xianghu@gxnu.edu.cn}
\affiliation{College of Physics and Technology, Guangxi Normal University, Guilin, 541004, China}
\author{Xuedong Tian}
\email{Contact author: snowtxd@gxnu.edu.cn}
\author{Liang Du}
\email{Contact author: liangdu@gxnu.edu.cn}
\affiliation{College of Physics and Technology, Guangxi Normal University, Guilin, 541004, China}
\begin{abstract}
We investigate the nonequilibrium dynamics of charge density waves in a pumped one-dimensional Hubbard superlattice with staggered onsite Coulomb interactions at half-filling, using time-dependent 
exact diagonalization. In equilibrium, the system exhibits commensurate charge correlations consistent with the superlattice periodicity. Under laser excitation, the charge correlation function exhibits distinct behaviors across four representative frequencies, spanning both linear and nonlinear optical regimes. 
Notably, we observe a laser-induced commensurate-to-incommensurate transition in the charge order, manifested by a shift in the peak wavevector of the charge structure factor. 
This transition is driven by sublattice-selective doublon-holon dynamics, where the laser frequency and intensity determine whether excitations predominantly destabilize the charge order on the weakly or strongly interacting sublattice. Our analysis of the excitation spectrum and site-resolved correlation dynamics reveals the underlying mechanisms of this transition. These results suggest a promising optical strategy for controlling charge order in superlattice-based quantum materials.
\end{abstract}
\date{\today}
\maketitle

\section{INTRODUCTION}
\label{sec:intro}
The non-equilibrium control of strongly correlated electronic systems using ultrafast optical techniques represents a vibrant frontier in modern condensed matter physics. This approach offers an unprecedented pathway to manipulate quantum states on ultrafast timescales and potentially create novel phases of matter with no equilibrium analogues \cite{aoki:rmp2014,delatorre:rmp2021,murakami:rmp2025}. 
By leveraging the mutual interplay between electrons, spins, and lattices, it allows for the dynamical engineering of material properties through selective excitation of specific degrees of freedom \cite{husq:cpl2023}. 
A prominent and widely studied example is the light-induced insulator-to-metal transition, where a transient metallic state is created by photo-doping or by coherently melting electronic orders \cite{rini:nat2007,caviglia:prl2012}. 
Beyond this, recent experiments have demonstrated a rich variety of photo-induced phenomena, including the enhancement of superconductivity\cite{fausti:sci2011}, the control of magnetic order\cite{mentink:nc2015}, and the emergence of hidden states such as  transient charge density waves\cite{kaneko:prl2019,luht:prl2012}. 
These advances highlight the potential of photoexcitation not only to probe but also to transiently stabilize and control phases that are inaccessible in thermal equilibrium, opening new avenues for understanding and harnessing strong correlations in quantum materials.

In low-dimensional correlated systems, the formation of a charge density wave typically necessitates additional ingredients, such as lattice modulation or long-range Coulomb interactions \cite{fabrizio:prl1999,tsuchiizu:prl2002,ejima:prl2007,Paiva:prb2002}. 
The introduction of a superlattice potential—mimicking heterostructures or periodically strained chains—explicitly breaks translational symmetry, thereby enabling a rich competition between spin density wave (SDW) and charge density wave (CDW) instabilities driven by electron-electron interactions. While equilibrium properties of such superlattice systems have been extensively studied, their non-equilibrium dynamics under external stimuli, particularly ultrafast laser excitation, remain far less explored \cite{luht:prl2012,cheng:prb2024,cheng:prl2025}. 
This gap in knowledge prompts several key questions: Can tailored laser pulses not only enhance or suppress charge order but also fundamentally alter its commensurability? 
Moreover, how do the CDW orders evolve dynamically, and what are the underlying microscopic processes - in terms of Fock states dynamics - when driven across the commensurate-incommensurate phase boundary?  Our work addresses these open questions by investigating the laser-induced dynamics of a correlated Hubbard superlattice.

The significance of such control is underscored by the fact that the commensurability of charge order and its instability are often invoked as a key factor in understanding the complex interplay—ranging from competition to coexistence—between charge density waves and superconductivity in various unconventional superconductors, such as cuprates \cite{arx:npjqm2023} and other systems \cite{Li2016, Kogar2017}. Thus, achieving non-equilibrium control over charge commensurability in a highly tunable superlattice platform, as demonstrated here, provides a novel pathway to simulate and dissect these competing mechanisms in a minimal and well-controlled setting.

In this work, we investigate a one-dimensional Hubbard superlattice at half-filling, 
characterized by a periodic alternation of on-site Coulomb interactions. 
This model serves as a fundamental theoretical framework for understanding correlated electron systems with intrinsic spatial inhomogeneity \cite{Paiva:prl1996,Paiva:prb1998,Paiva:prb2000,Paiva:prb2002,duan:jpcm2010,zhangll:cpb2015}. 
In a condensed matter context, such superlattices can be experimentally realized in materials with nanoscale heterogeneity, forming quasi-one-dimensional chains where atoms with differing orbital energies and electron correlation strengths alternate. 
Prominent physical realizations include the one-dimensional copper-oxide model \cite{Emery:prl1990,schlappa:nat2012,chenz:sci2021} and atomic chains composed of carbon and transition-metal compounds \cite{Dag:prb2005}. 
The simplest manifestation of this inhomogeneity is a two-site unit cell, with one site (A-sublattice) having an on-site Coulomb interaction $U_\mathrm{A}$ and the other (B-sublattice) characterized by $U_\mathrm{B}$.
The ground state properties of this system are highly sensitive to the interplay between $U_\mathrm{A}$ and $U_\mathrm{B}$. 
When the B sublattice is non-interacting ($U_\mathrm{B} = 0$) and $U_\mathrm{A} > 0$, the system retains particle-hole symmetry and surprisingly avoids a Mott insulating phase. 
Instead, it exhibits a correlated metallic state with gapless spin and charge excitations \cite{duan:jpcm2010,zhangll:cpb2015}. 
However, once a finite Coulomb interaction $U_\mathrm{B} > 0$ is introduced on the B sublattice, a qualitative transition occurs. The system enters an antiferromagnetic Mott insulating phase, characterized by a finite charge gap while spin excitations remain gapless, reflecting its underlying magnetic ordering \cite{duan:jpcm2010}. 
Crucially, fixing $U_\mathrm{A}$, and varying $U_\mathrm{B}$ can induce a Commensurate-to-Incommensurate (C-IC) transition in the charge correlations.  For instance, at $U_\mathrm{A}=18, U_\mathrm{B}=3.0$, the charge correlation is commensurate, characterized by the maximum momentum $q_\mathrm{max} = \pi$ in the electron structure factor $N(q)$ \cite{duan:jpcm2010}. This established equilibrium phenomenon sets the stage for our non-equilibrium investigation. 

We employ time-dependent exact diagonalization to investigate the photoinduced dynamics in an one-dimensional Hubbard superlattice with a staggered onsite Coulomb interaction at half-filling \cite{Mondaini:prb2017,cheng:prb2024}. 
Going beyond the previously reported selective enhancement of doublons on A or B sublattices depending on laser parameters \cite{cheng:prb2024}, we demonstrate that laser excitation can drive a dynamical transition from a commensurate to an incommensurate charge-ordered phase. 
This transition is identified by a shift of the peak wavevector in the charge structure factor 
$N(q,t)$ away from momentum $q_\mathrm{max} = \pi$. We systematically analyze 
this phenomenon under various laser frequencies, 
spanning both linear and nonlinear excitation regimes, and elucidate how the competition between inter- and intra-sublattice correlations drives this transition by analyzing site-resolved charge correlations and doublon-holon dynamics.
Our findings suggest a novel optical strategy for controlling the commensurability of charge order in Hubbard superlattices, 
with potential implications for the design of optically switchable quantum devices.

The remainder of this paper is organized as follows.
In Sec.\ref{sec:model}, we describe the Hamiltonian of the pumped one-dimensional Hubbard model with a modulated (staggered) site-dependent Coulomb interaction, and the time-dependent Lanczos method of solution.
In Sec.\ref{sec:eqset}, the equilibrium sublattice site-specific density of states and the charge correlation competetion between different sublattices are illustrated  using exact diagonalization.
In Sec.\ref{sec:neqdy}, we study the non-equilibrium transition between commensurate and incommensurate charge correlations, with detailed analysis.
Finally, in Sec.\ref{sec:concl} we present the main conclusions and discussions of the paper.

\section{model and method}
\label{sec:model}
We consider a one-dimensional Hubbard superlattice at half-filling, described by the following equilibrium Hamiltonian with particle-hole symmetry \cite{duan:jpcm2010,zhangll:cpb2015,Mondaini:prb2017},
\begin{align}
    H &= - \tsh \sum_{i\sigma} \left(c_{i\sigma}^\dagger c_{i+1\sigma}^{} + \mathrm{H.c.} \right) \nonumber\\
      &\quad + \sum_i U_{i}  (n_{i\uparrow} - \frac{1}{2}) (n_{i\downarrow} - \frac{1}{2}),
\label{eq:eqH}
\end{align}
where $c_{i\sigma}$ ($c_{i\sigma}^\dagger$)  annihilates (creates) an electron at site $i$ with spin projection
$\sigma = \uparrow, \downarrow$, and $n_{i\sigma} = c_{i\sigma}^\dagger c_{i\sigma}$ is the corresponding electron number operator. The hopping amplitude between nearest-neighbor
sites is denoted by $\tsh$. The on-site Coulomb interaction $U_{i}$ between spin-$\uparrow$ and $\downarrow$ electrons alternates along the chain: $U_{i} = U_\mathrm{A}$ for odd sites and $U_i = U_\mathrm{B}$ for even sites. In this work, we set $\tsh =1 $ as the energy unit and correspondingly, the unit of time is the inverse of energy, $\tsh^{-1}$. 

The site-specific local density of states is defined as,
\begin{align}
\rho(\omega) = \sum_{i,\sigma} \sum_{\phi} 
    &|\langle\phi|c_{i\sigma}^\dagger|\psi_0\rangle|^2 \delta(\omega - E_\phi + E_0)  \nonumber\\
   +& |\langle\phi|c_{i\sigma}       |\psi_0\rangle|^2 \delta(\omega + E_\phi - E_0),
   \label{eq:eqdos}
\end{align}
where $\{|\phi\rangle\}$ are the eigenstates of the equilibrium Hamiltonian in Eq.\eqref{eq:eqH} with energies $E_\phi$, and $|\psi_0\rangle$ is the ground state with energy $E_0$. 

The static charge correlation function is defined as,
\begin{align}
    C_i(r) = \langle\psi_0| n_{i}n_{i+r}|\psi_0\rangle - \langle\psi_0| n_{i}|\psi_0\rangle \langle\psi_0| n_{i+r}|\psi_0\rangle,
\end{align}
where the two sites are $i$ and $i+r$ with distance $r$. For example, $i\in \mathrm{A},i+r\in \mathrm{B}$ gives $C_\mathrm{AB}(r)$, $i\in \mathrm{A}, i+r\in \mathrm{A}$ gives $C_\mathrm{AA}(r)$, $i\in \mathrm{B}, i+r\in \mathrm{B}$ gives $C_\mathrm{BB}(r)$. Correspondingly, the static charge correlation structure factor is written as,
\begin{align}
    N(q) &= \frac{1}{N} \sum_i \sum_r e^{iqr} C_{i}(r) \nonumber\\
         &= \frac{1}{N} \sum_i \sum_{r>0} 2 C_{i}(r) \cos(qr) + C_i(0).
\end{align}
where the relation $C_i(r) = C_i(-r)$ has been used. 
To characterize the C-IC transition, we employ the following criterion derived from the second derivative of $N(q)$ \cite{zhangll:mplb2015}, 
\begin{align}
     \mathbb{F} = \frac{1}{N}\sum_{r>0,i}2 (-1)^{r+1} r^2 |C_i(r)|.
     \label{eq:cic}
\end{align}
A commensurate charge correlation state corresponds to $\mathbb{F} > 0$, while an incommensurate state corresponds to $\mathbb{F} < 0$.  

As the system is driven out of equilibrium, we simulate external laser pulses in a time gauge by means of a time-dependent vector potential $A(t)$ (directed along the chain direction)\cite{Hiroaki:jpsj2012},
\begin{align}
    A(t) = A_0 \exp[-(t-t_p)^2/2t_d^2] \cos[\Omega (t-t_p)],
\end{align}
where the laser pulse is characterized by its amplitude $A_0$, frequency $\Omega$, and a temporal envelope. The laser pulse is peaked at $t_p$, with $t_d$ characterizing the duration time (pulse width) of light. For our numerical simulations, we set $t_p = 8.0, t_d = 2.0$ in the following numerical calculations. 
The laser excitation is incorporated in the Hamiltonian through the Peierls substitution, which modifies the hopping term,
\begin{align}
    H &= - \tsh \sum_{i\sigma} \left[ e^{i A(t)} c_{i\sigma}^\dagger c_{i+1\sigma} + \mathrm{H.c.} \right] \nonumber\\
    &\quad +  \sum_i U_{i} (n_{i\uparrow} - \frac{1}{2}) (n_{i\downarrow} - \frac{1}{2}),
\label{peierls}
\end{align}
We adopt a chain size of $L=14$ and a coarse-grained time $\delta t = 0.005 \tsh^{-1}$. The system is set at half-filling, i.e., the number of electrons $N$ equals the number of sites $L$. Furthermore, we restrict the calculation to the subspace of zero total magnetization, ensuring $N_{\uparrow} = N_{\downarrow}$.

The numerical simulation proceeds as follows. First, the ground state $\psi(t=0^{-})$ of the initial Hamiltonian is computed using the Lanczos exact diagonalization method. This state is subsequently used to initialize the time-dependent Schrödinger equation, $i\partial_t |\Psi(t)\rangle = H(t) |\Psi(t)\rangle$.
The time evolution is implemented step-by-step based on the time-dependent Lanczos method \cite{Innerberger:epjp2020,luht:prl2012,Park:jcp1986,Mohankumar:cpc2006,Balzer:jpcm2012},
\begin{equation}
  |\Psi(t+\delta t)\rangle \approx e^{-i H(t)\delta t} |\Psi(t)\rangle
                           \approx \sum_{l=1}^M e^{-i \epsilon_l^{} \delta t} |\Phi_l\rangle\langle \Phi_l| \Psi(t)\rangle,\nonumber
\end{equation}
where $\epsilon_l^{}$ ($\Phi_l$) are the eigenvalues (eigenvectors) of the tri-diagonal matrix generated by Lanczos iteration with a Krylov subspace size of $M \leq 100$. (The required $M$ for a given accuracy depends on the setup and the chosen time step \cite{Moler:siamr2003}.)
We set the time-step size to $\delta t = 0.005t_h^{-1}$ in our calculation of the time evolution.
Physical observables are computed as the expectation value,
\begin{align}
     \langle O\rangle_t = \langle \Psi(t)| O |\Psi(t)\rangle.
\end{align}
The time-dependent charge correlation function is defined as,
\begin{align}
    C_i(r,t) = \langle n_{i}n_{i+r}\rangle_t - \langle n_{i}\rangle_t \langle n_{i+r}\rangle_t,
\end{align}
and the corresponding time-dependent charge structure factor is given by,
\begin{align}
    N(q,t) = \frac{1}{L}\sum_{i=1}^L\sum_{r=0}^{L-1}e^{iqr} C_i(r,t).
\end{align}
Note that the $r=0$ terms are independent of momentum $q$.

\section{Equilibrium Background and Non-equilibrium Setup}
\label{sec:eqset}
\begin{figure}[t]
    \centering
    \includegraphics[width=0.45\textwidth]{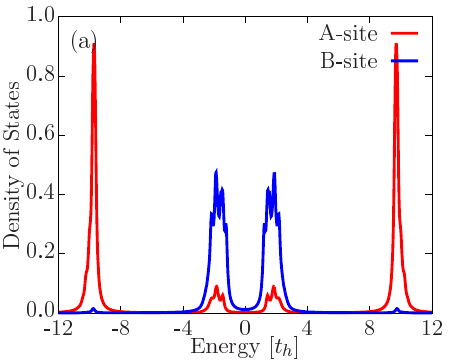} 
    \includegraphics[width=0.45\textwidth]{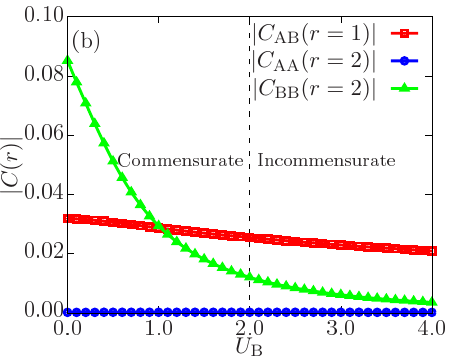} 
    \caption{(a) Sublattice resolved density of states, where lower and upper Hubbard bands and hybridization bands are labeled, and a broadening parameter $\eta = 0.1$ is employed.(b) For fixed $U_\mathrm{A}=18.0$, correlation functions as a function of $U_\mathrm{B}$ are plotted for AA, AB, and BB lattice sites, respectively. The critical $U_\mathrm{B} \approx 2.0$ between incommensurate-commensurate electron correlation is labeled as a dashed line.}
    \label{fig1} 
\end{figure}

To establish a physical context for our non-equilibrium study, we first investigate the ground-state properties of the one-dimensional Hubbard superlattice model at half-filling. 
We employ a chain of length $L=14$ with alternating on-site Coulomb interactions  $U_\mathrm{A}=18.0$ and $U_\mathrm{B}=3.0$. Fig.\ref{fig1}(a) shows the zero-temperature density of states (DOS) for this model. The spectrum exhibits particle-hole symmetry in both sublattices (A and B). 
A pronounced energy gap at the Fermi level ($E_{F} = 0.0$) identifies the ground state as an insulator. Notably, the DOS for each sublattice features four distinct peaks. 
For the A sublattice, the lower and upper Hubbard bands are located at approximately $\pm U_\mathrm{A}/2$, while two hybridization bands appear near $\pm U_\mathrm{B}/2$. Conversely, for the B sublattice, the Hubbard bands are located at roughly $\pm U_\mathrm{B}/2$, with weak hybridization features near $\pm U_\mathrm{A}/2$. This sublattice-resolved electronic structure is crucial for understanding the site-selective charge dynamics induced by laser excitation.

As outlined in the introduction, the competition between the staggered Coulomb interactions $U_\mathrm{A}$ and $U_\mathrm{B}$ gives rise to a rich phase diagram. 
A central feature of this diagram is the transition between commensurate and incommensurate charge-ordered phases. 
To map this out, we fix $U_\mathrm{A} = 18.0$ to model a strongly correlated site and systematically vary $U_\mathrm{B}$, from $0.0$ to $4.0$. 
The nature of the emerging charge order is characterized by the peak wavevector $q_\mathrm{max}$ of the static charge structure factor $N(q)$ in the ground state. 
Here, a commensurate charge density wave (CDW) is identified by $q_\mathrm{max} = \pi$, while a deviation from $\pi$ signifies an incommensurate CDW.

The observed behavior can be understood as follows: At  
$U_\mathrm{B} = 0$, the system is a correlated metal with incommensurate charge fluctuations, where next-nearest-neighbour (NNN) correlation (primarily between B sites) dominate over nearest-neighbour (NN) (AB) correlation \cite{duan:jpcm2010}. As 
$U_\mathrm{B}$ increases, the charge order on the B-sublattice strengthens and eventually locks in phase with that on the A-sublattice. 
This leads to an incommensurate-to-commensurate transition at a critical value  
$U_\mathrm{B}^c \approx 2.0$. 
For $U_\mathrm{B} > U_\mathrm{B}^c$, the charge order is fully commensurate, characterized by a perfect staggering of charge correlation between the A and B sublattices, with $q_\mathrm{max} = \pi$. 

This transition is quantitatively captured by the evolution of the correlation functions in Fig.\ref{fig1}(b). In the incommensurate phase ($U_\mathrm{B} < U_\mathrm{B}^c$), the next-nearest-neighbor correlation
$C_\mathrm{BB}(r=2)$ is significant. As $U_\mathrm{B}$ increases, $C_\mathrm{BB}(r=2)$ is strongly suppressed,
whereas the dominant nearest-neighbor inter-sublattice correlation $C_\mathrm{AB}(r=1)$ remains relatively robust. 
The commensurate condition, derived from the criterion $\mathbb{F} > 0$ in Eq.\eqref{eq:cic}, requires the inter-sublattice correlation to dominate over the intra-sublattice contributions, which in the strong coupling limit simplifies to $|C_\mathrm{AB}| > 2 |C_\mathrm{BB}|$  \cite{zhangll:cpb2015,zhangll:mplb2015}. The crossing point of
$|C_\mathrm{AB}|$ and $2 |C_\mathrm{BB}|$, marked by the dashed line in Fig.\ref{fig1}(b),  
approximately coincides with the critical $U_\mathrm{B}^c = 2.04$, 
thereby confirming that the in-equilibrium C-IC transition is driven by 
the competition between diminishing B-sublattice correlations and 
prevailing A-B inter-sublattice correlations.

Based on this equilibrium scan, we select $U_\mathrm{B}=3.0$ as the working point for all subsequent time-dependent simulations, a parameter at which the system exhibits commensurate charge order. 
This choice is strategic for two key reasons:
(1) It places the initial state deep within the commensurate charge-ordered phase, providing a clean and well-defined starting point.
(2) It allows us to explore a fundamental question: Can a tailored laser pulse drive the system dynamically from this well-established commensurate phase into an incommensurate one, thereby effectively reversing the equilibrium phase transition?
This sets the stage for the non-equilibrium results presented in the following sections, where we apply laser pulses of various frequencies to this commensurate ground state. 
\begin{figure}[t]
    \centering
    \includegraphics[width=0.45\textwidth]{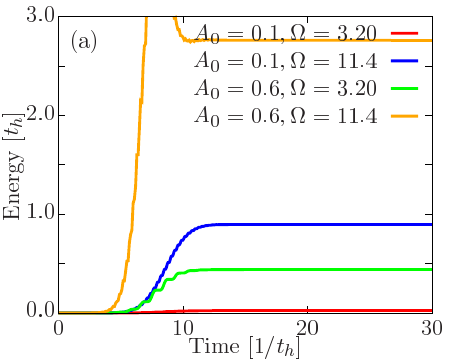}
    \includegraphics[width=0.45\textwidth]{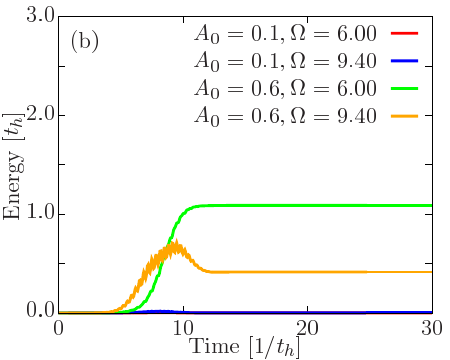}
    \caption{Time evolution of the total energy as the studied system is driven by laser. Two represented laser intensity $A_0 = 0.1, 0.6$ are adopted, respectively. Note, a shift of ground state energy to zero is contained. (a) Single photon processes with laser frequency $\Omega = 3.2,11.4$, (b) Multi-photon processes with $\Omega = 6.0, 9.4$.}
    \label{fig2} 
\end{figure}

Guided by our earlier out-of-equilibrium study of the Hubbard superlattice \cite{cheng:prb2024}, we strategically choose laser frequencies based on the energy absorption spectrum. 
We select $\Omega = 3.2$ and $\Omega = 11.4$ to resonantly target the two dominant single-photon absorption peaks. 
These frequencies correspond to distinct inter-sublattice charge excitation channels: $\Omega = 3.2$ primarily promotes electrons from the lower Hubbard band of the B-sublattice to the upper hybridization band of the A-sublattice (or the reverse process), while $\Omega = 11.4$ drives higher-energy transitions between the Hubbard bands of the two sublattices.
To explore the nonlinear optical regime, we employ frequencies of $\Omega = 6.0$ and $\Omega = 9.4$, which are associated with two-photon absorption processes \cite{cheng:prb2024}. This selection enables a comprehensive study of the system's response across both linear and nonlinear excitation regimes. 

The distinction between these processes is directly reflected in our choice of laser intensity.
For the single-photon processes ($\Omega = 3.2, 11.4$), a relatively weak laser intensity ($A_0 = 0.1$) is sufficient to drive significant dynamics, as the system is directly excited via a resonant, dipole-allowed pathway. Conversely, for the two-photon processes ($\Omega = 6.0,9.4$), a stronger intensity ($A_0=0.6$) is required to observe notable effects, as multi-photon absorption is a nonlinear process whose probability scales with a higher power of the field amplitude. The energy absorption dynamics are plotted in fig.\ref{fig2}(a). We note that the energy after the pulse remains constant due to the unitary nature of the time evolution (no energy is added or removed from the system).

The laser driven energy absorption are plotted in Fig.\ref{fig2}(b). At the weak intensity ($A_0=0.1$), the system exhibits negligible energy absorption and dynamic response for these off-resonant frequencies, as the perturbative linear response is ineffective. In contrast, at the stronger intensity ($A_0 = 0.6$), the increased field strength efficiently drives the system via these virtual, nonlinear pathways, leading to substantial energy absorption and the distinct charge dynamics that will be detailed in the subsequent sections. This deliberate parameter selection thus enables a comprehensive study spanning both the linear and nonlinear excitation regimes.
\begin{figure*}[ht]
\centering
\includegraphics[angle=-0,width=0.245\textwidth]{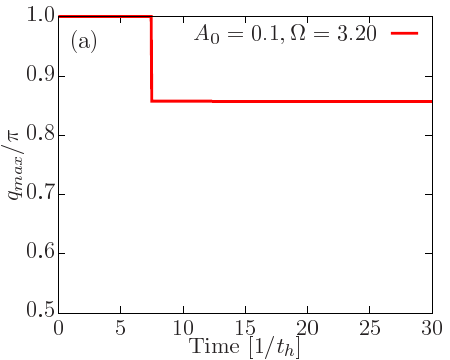}
\includegraphics[angle=-0,width=0.245\textwidth]{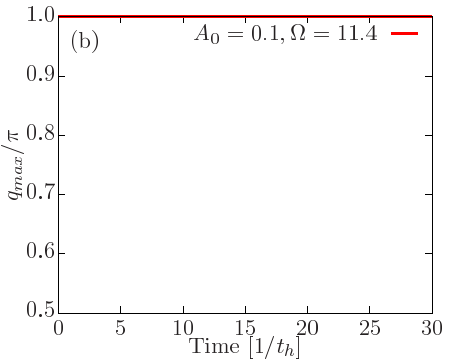}
\includegraphics[angle=-0,width=0.245\textwidth]{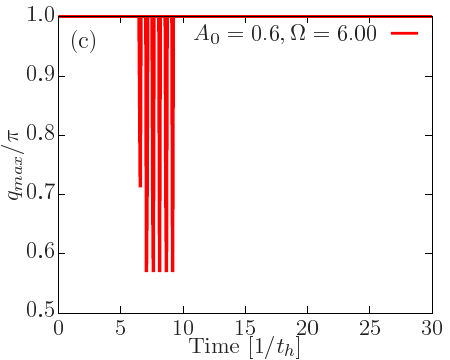}
\includegraphics[angle=-0,width=0.245\textwidth]{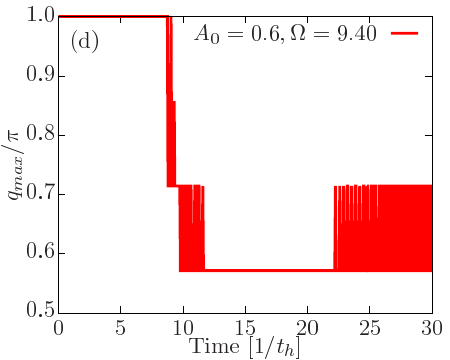}
\includegraphics[angle=-0,width=0.245\textwidth]{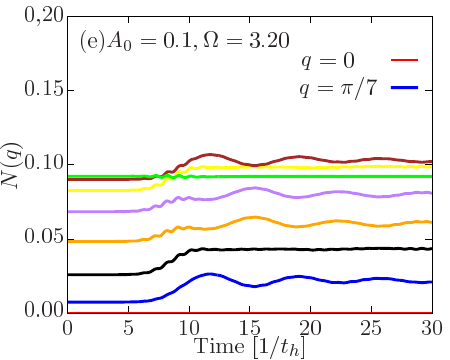}
\includegraphics[angle=-0,width=0.245\textwidth]{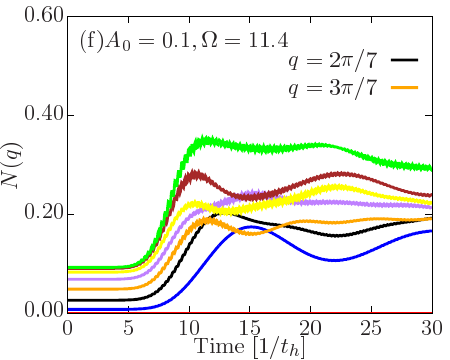}
\includegraphics[angle=-0,width=0.245\textwidth]{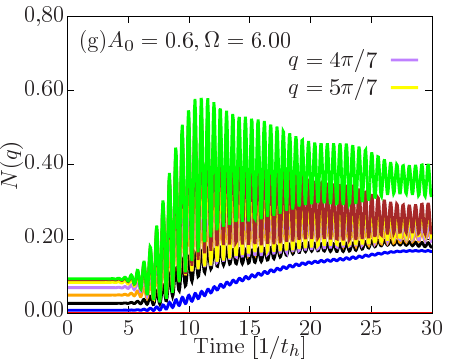}
\includegraphics[angle=-0,width=0.245\textwidth]{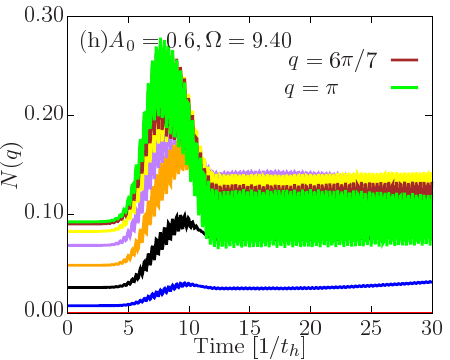}
\caption{Laser-induced commensurate-incommensurate transition dynamics. (a-d) Time evolution of the normalized peak wavevector $q_\mathrm{max}(t)/\pi$ 
of the charge structure factor for different laser frequencies. (e-h) Corresponding momentum- and time-resolved evolution of the charge structure factor $N(q,t)$. The system parameters are  $L=14$, $U_\mathrm{A} = 18.0$ and $U_\mathrm{B} = 3.0$ with laser parameters: (a,e) $\Omega = 3.2, A_0 = 0.1$; (b,f) $\Omega = 11.4,  A_0 = 0.1$; (c,g) $\Omega = 6.0,  A_0 = 0.6$, (d) $\Omega = 9.4,  A_0 = 0.6$.}
\label{fig3}
\end{figure*}
\section{Laser-Induced Dynamical Commensurate-to-Incommensurate Transition}
\label{sec:neqdy}
Having established the equilibrium phase diagram and the non-equilibrium setup, we now present the central result of this work: the laser-induced dynamical commensurate-to-incommensurate (C-IC) transition of charge order. We monitor this transition in real time through the evolution of the charge structure factor $N(q,t)$, specifically by tracking the wavevector $q_\mathrm{max}(t)$  at which $N(q,t)$ attains its maximum value. In the initial commensurate phase, $q_\mathrm{max}(0) = \pi$. 
A sustained dynamical shift of $q_\mathrm{max}(t)$ away from $\pi$ signals the emergence of incommensurate charge correlations. 

The corresponding dynamics of the key order parameter—the normalized peak wavevector $q_\mathrm{max}(t)/\pi$ are quantified in Figure \ref{fig3}. 
The top row of panels (a-d) shows the time evolution of $q_\mathrm{max}(t)/\pi$ for the four laser frequencies, while the bottom row (e-h) displays the corresponding full momentum- and time-resolved charge structure factor $N(q,t)$.
The system's response is highly frequency-selective, revealing distinct pathways for manipulating charge order.

At the low single-photon frequency $\Omega=3.2$ (Fig.\ref{fig3}(a)), the laser pulse promptly induces a shift in the charge structure factor peak. 
The value of $q_\mathrm{max}(t)/\pi$ drops below 1.0 and stabilizes at an incommensurate value (6/7) after the pulse, indicating a stable dynamical transition to an incommensurate phase. 
At the high single-photon frequency $\Omega=11.4$ (Fig.\ref{fig3}(b)), the charge structure factor remains largely unchanged, with $q_\mathrm{max}(t)$ firmly pinned at $\pi$, 
indicating that this excitation pathway preserves the commensurate order. 
For the two-photon process at $\Omega=6.0$ (Fig.\ref{fig3}(c)), the effect is transient during the laser pulse. 
A slight deviation from commensurability occurs only during the intense part of the laser pulse (centered at $t_p=8$), 
after which the system rapidly relaxes back to the commensurate state $q_\mathrm{max}(t)/\pi = 1)$. 
In contrast, the two-photon process at $\Omega=9.4$ (Fig.\ref{fig3}(d)) drives the most pronounced C-IC transition. The suppression of $q_\mathrm{max}(t)/\pi$ is stronger (oscillate between $5/7$ and $4/7$) than in the $\Omega = 3.2$ case, demonstrating that nonlinear excitation can be highly effective in destabilizing the commensurate order.

The rich dynamics of the C-IC transition are further visualized in the momentum- and time-resolved evolution of the full charge structure factor $N(q,t)$, shown in Fig.\ref{fig3}(e-h). In the initial equilibrium state, $N(q,t=0)$  increase monotonically from $q=0$ to $q=\pi$, with a single sharp peak at $q=\pi$ characterizing the initial commensurate charge order. The application of the laser pulse dramatically alters this monotonic profile, inducing distinct non-monotonic structures that signal the transformation of charge order. In Fig.\ref{fig3}(e) ($\Omega = 3.2$) and Fig.\ref{fig3}(h) ($\Omega = 9.4$), the primary peak visibly shifts and splits away from $q=\pi$ during and after the pulse. A stable incommensurate order is established post-pulse, evidenced by the main peak settling at a wavevector less than $\pi$ (e.g., at $q/\pi=6/7$ in Fig.\ref{fig3}(e)) and the concomitant emergence of a new shoulder or a secondary hump at a different incommensurate wavevector (e.g., near $q/\pi= 4/7, 5/7$). In stark contrast, for $\Omega = 11.4$ in Fig.\ref{fig3}(f), the commensurate order remains robust. While the overall intensity of $N(q,t)$  is enhanced by the pulse, the spectral weight remains overwhelmingly concentrated at $q=\pi$, with no significant shift of the peak maximum or development of new incommensurate features. 
For the  two-photon process at $\Omega = 6.0$ in Fig.\ref{fig3}(g), the response is transient. A slight broadening of the main peak and weak non-monotonic modulations appear only during the intense part of the laser pulse. However, these features vanish immediately after the pulse, and the structure factor fully reverts to its original monotonic profile with a single peak at $q=\pi$, consistent with the transient deviation observed in Fig.\ref{fig3}(c).The appearance of these non-monotonic features and the sustained displacement of the primary peak in 
$N(q,t)$ provide direct and unambiguous evidence of the laser-induced destabilization of the commensurate order and the formation of a dynamically stabilized incommensurate phase.
\begin{figure*}[htp]
    \centering
    \includegraphics[angle=-0,width=0.245\textwidth]{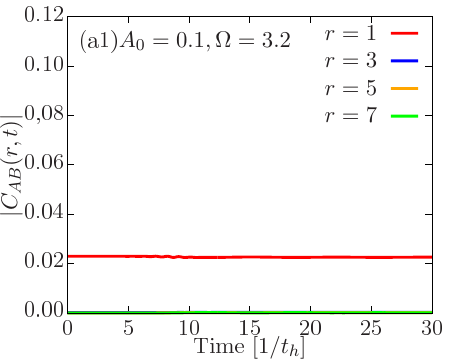}
    \includegraphics[angle=-0,width=0.245\textwidth]{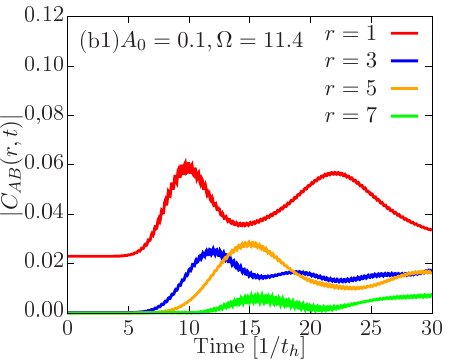}
    \includegraphics[angle=-0,width=0.245\textwidth]{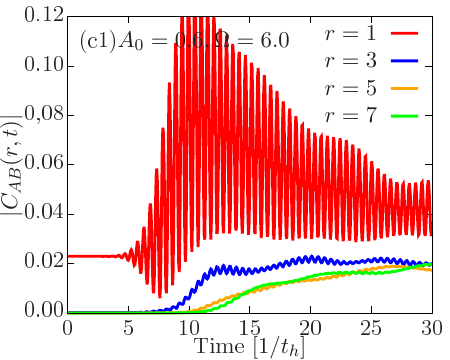}
    \includegraphics[angle=-0,width=0.245\textwidth]{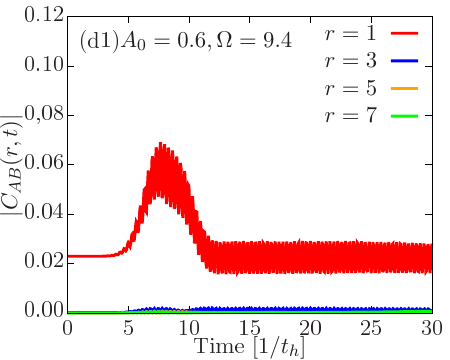}
    \includegraphics[angle=-0,width=0.245\textwidth]{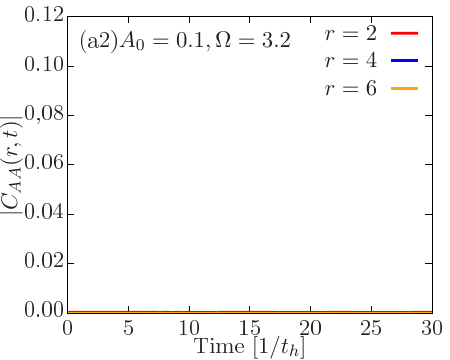}
    \includegraphics[angle=-0,width=0.245\textwidth]{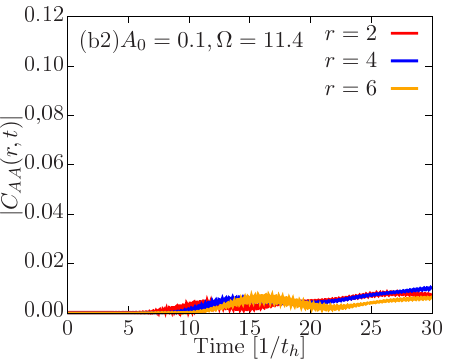}
    \includegraphics[angle=-0,width=0.245\textwidth]{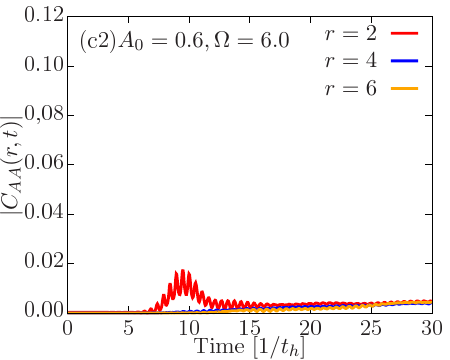}
    \includegraphics[angle=-0,width=0.245\textwidth]{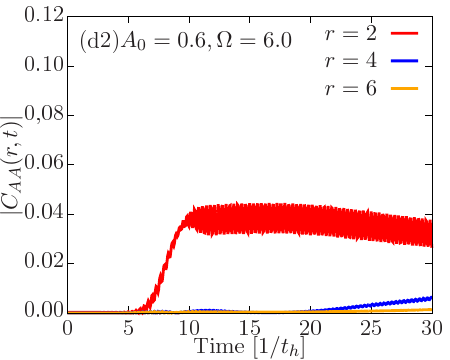}
    \includegraphics[angle=-0,width=0.245\textwidth]{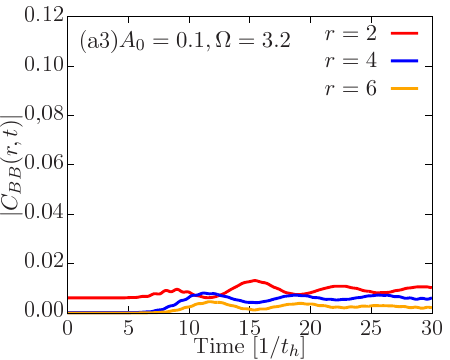}
    \includegraphics[angle=-0,width=0.245\textwidth]{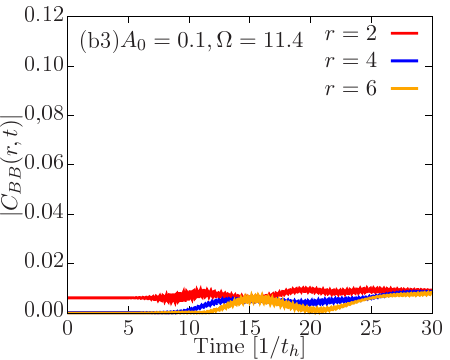}
    \includegraphics[angle=-0,width=0.245\textwidth]{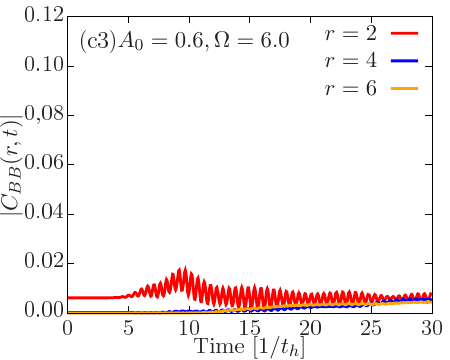}
    \includegraphics[angle=-0,width=0.245\textwidth]{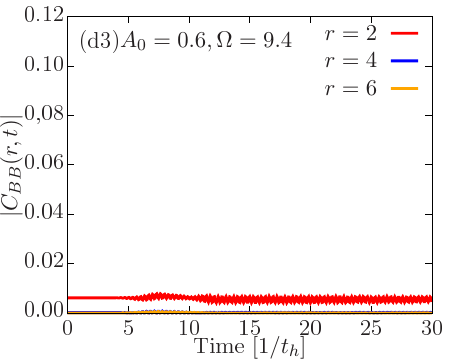}
\caption{Time evolution of the charge correlation function $C_{AA}, C_{AB}, C_{BB}$ for Hubbard superlattices with fixed position-dependent Coulomb interaction strengths $U_\mathrm{B} = 18.0$ and $U_\mathrm{B} = 3.0$ with chain size $L=14$, where $C_{AA}, C_{AB}, C_{BB}$ represent the A-A, B-B, A-B sublattice charge correlation function. (a) The single photon process with  $A_0=0.1, \Omega = 3.2$,
(b) The single photon process with  for $\Omega = 6.0$,
(c) The double photon process with  for $\Omega = 9.4$,
(d) The double photon process with  for $\Omega = 11.4$.}
\label{fig4}
\end{figure*}
To unravel the microscopic origin of these distinct dynamical responses, we analyze the site-resolved, time-dependent charge correlation functions $C_\mathrm{AA}(r,t)$
$C_\mathrm{BB}(r,t)$, and 
$C_\mathrm{AB}(r,t)$. 
These functions are defined by averaging over the respective sublattice pairs as follows:
\begin{align}
    C_\mathrm{AA}(r,t) &=  \frac{2}{L}\sum_{i \in A} C_i(r,t), \\
    C_\mathrm{BB}(r,t) &= \frac{2}{L} \sum_{i \in B} C_i(r,t), \\
    C_\mathrm{AB}(r,t) &= \frac{2}{L} \sum_{i \in A}C_i(r,t).
    \label{eq:corr}
\end{align}

The time evolution of these correlations under laser excitation is shown in Fig. \ref{fig4}. To quantitatively connect their behavior to the C–IC transition, we introduce a time-dependent generalization of the equilibrium criterion \cite{zhangll:mplb2015},
\begin{align}
     \mathbb{F} = \sum_{r>0}  r^2 \left(|C_\mathrm{AB}(r,t)| -  |C_\mathrm{AA}(r,t)| - |C_\mathrm{BB}(r,t)|\right),
     \label{eq:cict}
\end{align}
which captures the competition among the three distinct correlation channels.

In the initial equilibrium commensurate phase, the inter-sublattice correlation 
$C_\mathrm{AB}(r=1)$ dominates. The laser-induced dynamics can be understood as a competition that disrupts this balance. 
At $\Omega = 3.2$ (Fig.\ref{fig4}(a1-a3)), $C_\mathrm{AB}$ remains largely unchanged, while $C_\mathrm{AA}$ stays near zero due to the strong on-site Coulomb interaction and the relative small photon frequency. $C_\mathrm{BB}$, however, is significantly enhanced due to the weaker interaction on the B-sublattice. 
The localization of charge fluctuations on the B-sublattice, signified by this marked enhancement of $C_\mathrm{BB}$, breaks the perfect A-B staggering. This redistribution of correlations drives the system into an incommensurate state, which is quantified by the sign change of $\mathbb{F}(t)$ in Eq.\eqref{eq:cict}.
Furthermore, the non-equilibrium enhancement of longer-range correlations such as
$C_\mathrm{BB}(r=4)$ and $C_\mathrm{BB}(r=6)$ plays a crucial role due to the $r^2$ weighting in Eq.\eqref{eq:cict}. Analysis of the Fock states in the final excited state $|\psi(t=30)\rangle$ reveals that while spins on neighboring A sites maintain antiferromagnetic alignment, doublon-holon pairs form predominantly on adjacent B sites ($r=2$), giving rise to the observed enhancement in $C_\mathrm{BB}(r)$. For example, $|\uparrow,\uparrow\downarrow,\downarrow,0,...\rangle$ and $|\uparrow,\uparrow\downarrow,\downarrow,\uparrow,\downarrow,0,...\rangle$.

At the relative high laser frequency $\Omega = 11.4$ (Fig.\ref{fig4}(b1-b3)), all three correlations increase in a concerted manner, preserving the relative dominance of  
$C_{AB}$ and thus maintaining the initial commensurate state. Fock state analysis shows a significant enhancement of doublon-holon pairs, such as   $|\uparrow\downarrow,0,\uparrow,\downarrow,\uparrow,\downarrow,...\rangle$ and $|0,\uparrow\downarrow,\downarrow,\uparrow,\dots\rangle$, which lead to an increase in double-holon pair in neighboring A-B sites (an exciton) \cite{jeckelmann:prb2003} and as a result increase the A-B sublattice charge correlation.
At $\Omega = 6.0$ (Fig.\ref{fig4}(c1-c3)), the brief perturbation is associated with a moderate enhancement of 
$C_\mathrm{AB}$ itself, which slightly disrupts the commensurate balance only during the pulse. This can be understood as a two-photon process effectively generating a similar excitation profile to $\Omega =  11.4$ case, albeit with a weaker and transient impact. 

Finally, at $\Omega = 9.4$ (Fig.\ref{fig4}(d1-d3)), a different mechanism is at play. Here, the C-IC transition is driven by a strong enhancement of the intra-sublattice correlation 
$C_\mathrm{AA}$ on the strongly interacting A-sites. The competition between a now-dominant 
$C_\mathrm{AA}$ and the previously dominant 
$C_\mathrm{AB}$ results in the more pronounced incommensurate order observed in Fig.\ref{fig3}(d).

In summary, the observed trends in charge correlations are a direct manifestation of site-selective doublon generation triggered by the laser pulse. 
The laser frequency determines which sublattice is preferentially excited, leading to the formation of doublon-holon pairs at specific sites. The redistribution of these excitations alters the local charge environment, which in turn modifies the long-range correlation pattern, thereby either reinforcing or destabilizing the commensurate order.

Our results can be consistently understood through the lens of doublon formation: at $\Omega = 3.2$, doublons are primarily generated on the B-sublattice, consistent with the enhancement of 
$C_\mathrm{BB}$, whereas at  $\Omega = 9.4$, doublons are predominantly formed on the A-sublattice, leading to the increase of $C_\mathrm{AA}$. Furthermore, we have also studied the reverse process. When the initial state is situated in the incommensurate correlation regime (e.g., $U_\mathrm{A}=18.0, U_\mathrm{B}=1.5$), an incommensurate-to-commensurate transition is observed upon exposure to a characteristic high-frequency laser (data not shown). This demonstrates the bidirectional optical control of charge commensurability in our system.

\section{Conclusions and Discussions}
\label{sec:concl}
In summary, we have demonstrated a laser-induced commensurate-to-incommensurate transition of charge order in a one-dimensional Hubbard superlattice with staggered interactions. By employing time-dependent exact diagonalization, we show that this dynamical transition is driven by distinct, site-selective mechanisms: low-intensity, single-photon excitations primarily disrupt charge order via doublon formation on the weakly interacting sublattice, while high-intensity, multi-photon processes can more effectively destabilize it via excitations on the strongly interacting sublattice. The transition is quantitatively identified by a sustained shift of the peak wavevector in the time-dependent charge structure factor $N(q,t)$ away from $q_\mathrm{max} = \pi$.

Furthermore,, our work establishes that the commensurability of charge order—a fundamental property often considered static—can be dynamically controlled on ultrafast timescales.
Our findings highlight the efficacy of frequency-tuned laser pulses as a powerful tool for steering correlated electron systems into targeted states. 
The microscopic understanding of the competing inter- and intra-sublattice correlations provided here offers a new perspective on manipulating electronic phases in inhomogeneous correlated materials. 
The observation of the bidirectional transition (between commensurate and incommensurate phases) underscores the potential for all-optical control over quantum matter. We envision that such optical strategies could be applied to engineer functional non-equilibrium states in a wider class of superlattice-based quantum materials, paving the way for novel, optically switchable electronic and quantum devices. 
Beyond these specific findings,, and as discussed in the introduction, our work suggests a potential dynamical pathway to suppress commensurate charge order via tailored laser excitation, which could, in related materials, create a favorable backdrop for the emergence or enhancement of superconductivity. While this study focuses on the charge sector, the paradigm of using photo-modulation of correlations to manipulate competing orders offers fresh inspiration for exploring non-equilibrium quantum phenomena, including photo-induced superconductivity.

\acknowledgements
We acknowledge helpful discussions with Xiaojun Zheng and Cheng-Bo Duan. We gratefully acknowledge funding from the National Natural Science Foundation of China (Grant No. 12464018, and No. 12364022) 
and the Guangxi Young Elite Scientist Sponsorship Program (GXYESS2025075).
\bibliography{phmref}
\end{document}